\documentclass{mem}
\usepackage{natbib}\usepackage{txfonts}\usepackage{balance}
\usepackage{graphicx}
\usepackage[a4paper]{hyperref}
\idline{75}{282}
\begin{document}
\def\teff{$T\rm_{eff }$}
\def\kms{$\mathrm {km s}^{-1}$}

\title{
Present and future of the TeV astronomy with Cherenkov telescopes
}

   \subtitle{}

\author{
B. \,Sacco\inst{1} 
\and S. \,Vercellone\inst{1}
          }

  \offprints{B. Sacco}

\institute{
Istituto Nazionale di Astrofisica --
Istituto di Astrofisica Spaziale e Fisica Cosmica, Via U. La Malfa 153,
I-90146 Palermo, Italy
\email{bruno.sacco@iasf-palermo.inaf.it}
}

\authorrunning{Sacco \& Vercellone}

\titlerunning{TeV astronomy}

\abstract{
Cherenkov telescopes play a major role in the growth of the TeV Astronomy which, in 20 years, has reached 
the status of an important branch of Astrophysics, because of the observations of the violent, 
non thermal processes in the extreme band of the electromagnetic 
spectrum above several tens of GeV up to several tens of TeV. 
About one hundred extragalactic sources (Active Galactic Nuclei, blazars, and radiogalaxies) 
and Galactic sources (shell supernovae remnants, pulsar wind nebulae, isolated pulsars, X-ray binaries, 
and unidentified sources) have been detected so far. 

In the near future, an ambitious new array, the Cherenkov 
Compton Telescope (CTA) will substitute the present Cherenkov telescopes arrays.  
CTA is designed as an array of  many (50--100)  Cherenkov telescopes operated in stereo mode.  
CTA will allow to gain a factor of 10 in sensitivity with respect to the present arrays  such as
H.E.S.S., MAGIC, and VERITAS.
Moreover, CTA will  connect the TeV to the GeV energy band covered by
space missions such as Fermi and AGILE, and will also explore the highest energy region of the electromagnetic 
spectrum up to several  hundreds of TeV.

}
\maketitle{}

\section{Introduction: The role of the Cherenkov telescopes in the TeV astronomy}

\begin{figure*}[t!]
\resizebox{\hsize}{!}{\includegraphics[clip=true]{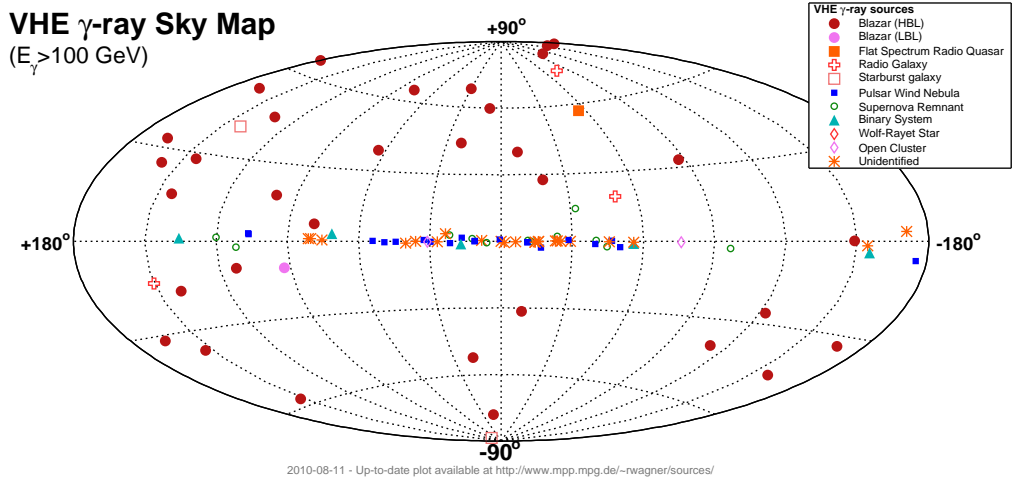}}
\caption{\footnotesize
TeV sources sky map (Energy $>$ 0.1 TeV).  From
{\tt http://www.mppmu.mpg.de/$\sim$rwagner/sources/index.html}
}
\label{fig:TeVsky}
\end{figure*}

\begin{figure*}
\resizebox{\hsize}{!}{\includegraphics[clip=true]{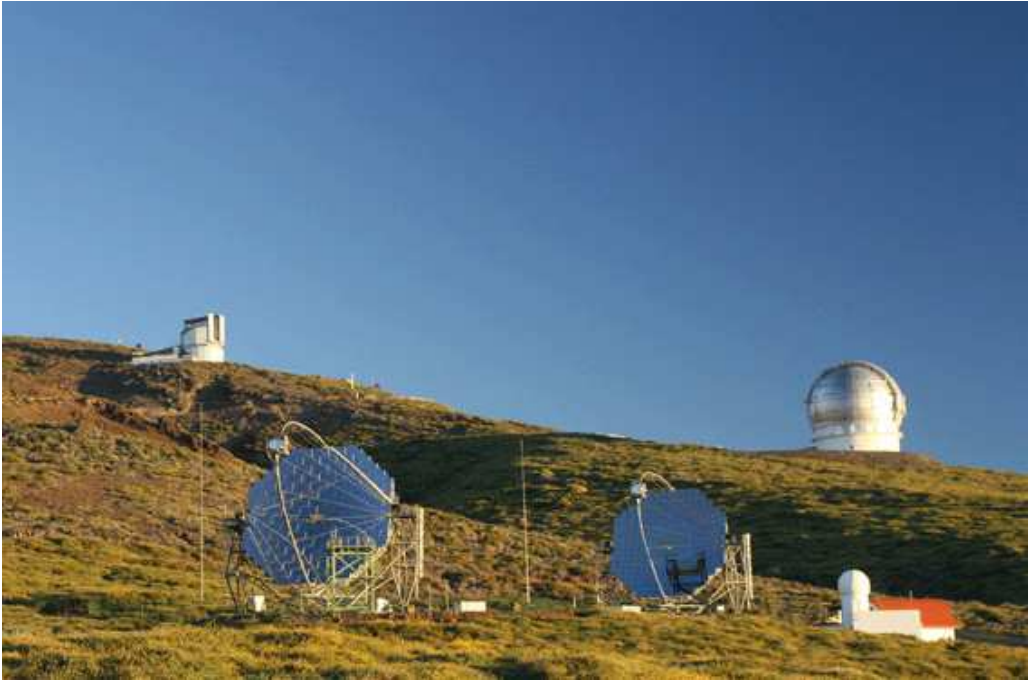}}
\caption{\footnotesize
A picture of the MAGIC array consisting of two  17\,m diameter Cherenkov telescopes,  in the foreground, 
the TNG  and GranTeCan telescopes in the
background on the left and on the right, respectively. 
They are located at Roque de los Muchachos, 2200 m a.s.l., La Palma , Canary Island. 
}
\label{fig:magic}
\end{figure*}

TeV astronomy deals with the observation of the emission from celestial objects from several 
tens of GeV up to hundreds of TeV. TeV astronomy was born in 1989, when the Whipple group 
reported  the observation of TeV $\gamma$-rays from the Crab Nebula using the atmospheric 
Cherenkov imaging technique \citep{Weekes1989}.  
Twenty years later, about 100 TeV sources have been discovered. Figure \ref{fig:TeVsky} shows a sky map of the 
TeV sources where both extragalactic and galactic objects belonging to different 
astrophysical classes are present: blazars, radio-galaxies, pulsar wind nebulae (PWN), supernova remnants, 
binary systems, and unidentified sources. 
The experiments which allowed this remarkable growth of the TeV astronomy belong to two main 
categories: 1) particles detectors, which measure, at the detector level, the particles of electromagnetic 
showers originated by an individual high energy $\gamma$-ray; and 2) Cherenkov telescopes, which measure 
the interaction of the electromagnetic cascade with the terrestrial atmosphere, imaging the lateral 
distribution of the charged relativistic particles. With respect to the particle detectors, Cherenkov telescopes 
reach a better performance in the $\gamma$/hadrons discrimination power, in the low energy threshold, 
and in the  angular resolution which can reach few arc-minutes when used in arrays. 
Due to this improved sensitivity, Cherenkov telescopes gave the major contribution to the TeV 
Astronomy development.
Table \ref{tab:telescopes} shows the main characteristics of the three most sensitive 
Cherenkov telescope arrays operating today in the world.
H.E.S.S. and VERITAS, in the southern and northern hemisphere, respectively, have similar characteristics, since
they operate in the energy range between 100 GeV and 10 TeV. The excellent sensitivity reached by 
H.E.S.S. allowed the discovery of the majority of the TeV sources known up to now. On the other hand, the main 
characteristics of MAGIC is its 17-m light-collector dish which allows to reach very low energy threshold, 
nearly at the boundary of the upper energy range of the $\gamma$-ray spectra collected by AGILE and Fermi space 
missions.
Therefore, the best performance of MAGIC is in the observation and study of 
AGNs and pulsars where the emission is more concentrated in the low energy range due to the  
extragalactic background light (EBL) effect for the AGNs and to the spectra cut-off for the pulsars.
\begin{table*}
\caption{Main characteristics of the three most sensitive Cherenkov telescope arrays.
$^{\star}$25 GeV with special triggers.}
\label{tab:telescopes}
\begin{center}
\begin{tabular}{lccc}
\hline
\\
 & \bf{H.E.S.S.} & \bf{MAGIC} & \bf{VERITAS} \\
\hline
\\
Site						& Namibia	& Canary Island	& Arizona\\
Latitude [$^{\circ}$]			& $-$25		& $+$29			& $+$32 \\
Height [km] 				& 1.8			&  2.2 			&1.3\\
N. of telescopes 			&   4			&  2				& 4\\
Telescope diameter [m]		& 12			& 17				& 12\\
Field of View [$^{\circ}$]		& 5.0			& 3.5				& 3.5\\
Low energy threshold [GeV]	& 100		& 50$^{\star}$		& 100\\
Sensitivity [mCrab]			& 7			& $<10$			& 10\\
\\
\hline
\end{tabular}
\end{center}
\end{table*}
Figure \ref{fig:magic} shows the MAGIC telescopes.
%
%

\section{TeV Sources}
%
%
\subsection{Extragalactic sources} 
More than 1/3 of the TeV sources are Active Galactic Nuclei (AGNs), mainly Blazars together with  
several radio-galaxies. 
In several blazars, the sub-TeV high energy  $\gamma$-ray emission is coupled to a strong emission at lower energy
(X-ray or UV). The strong temporal correlation between the TeV and the X-ray energy bands suggests a unique 
parent population for electrons emitting both inverse Compton in the very high energy $\gamma$-rays and  
synchrotron radiation in the X-ray and UV. The electrons are accelerated along a relativistic jet pointing towards
the observer \citep{Tavecchio1998}. Altogether, the unusual lack of temporal correlation found in some 
AGNs \citep[e.g. 1ES~1959$+$650,][]{Krawczynski2004:orphan} supports for the alternative hadronic 
models for the origin of the TeV emission \citep{Mucke2003:hadron}.  

\begin{figure}
\resizebox{\hsize}{!}{\includegraphics[clip=true]{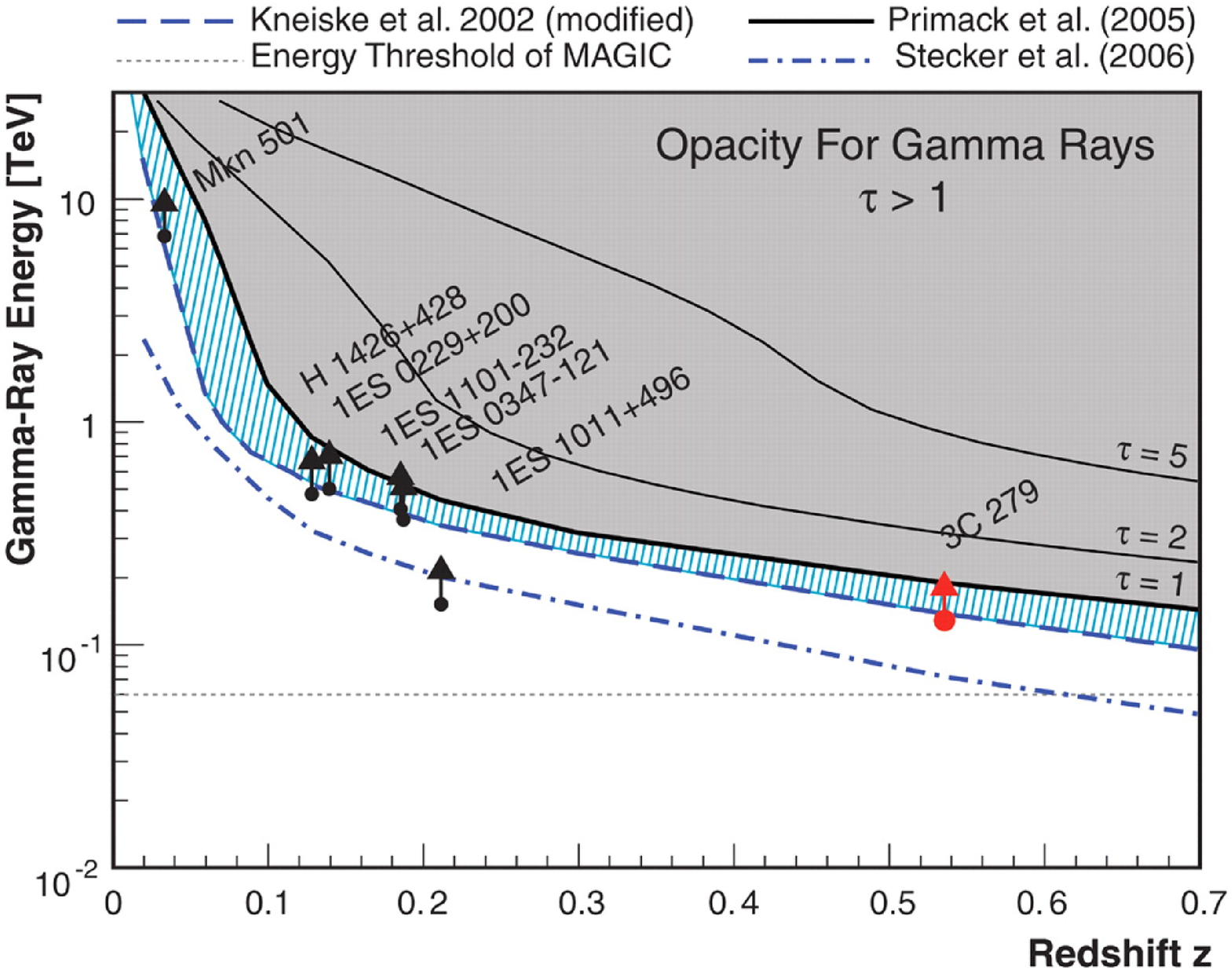}}
\caption{\footnotesize
The VHE $\gamma$-ray horizon. From \cite{Albert2008:3c279}.
}
\label{fig:3c279}
\end{figure}

The presence of the extragalactic background light confines the TeV AGN visibility. 
For this reason, the detection of 3C~279 by MAGIC \citep{Albert2008:3c279} puts severe
constraints on the opacity of the Universe, as shown in Figure \ref{fig:3c279}. 
Excluding 3C~279, whose redshift is $z=0.5362$, typical z values
are spread in the interval 0.0--0.2. Constraints on the EBL intensity can be derived by  the AGN spectral 
cut-off assuming an intrinsic AGN emission spectrum \citep{Raue2009}. 
Systematic errors in the EBL limit derive by the intrinsic spectral shape assumed and only a greater number 
of  blazars detected at redshift higher than 0.5 can provide more reliable constraints to the EBL.
Despite the fact that the EBL also limits the detection of very high energy $\gamma$-rays from GRBs,
several efforts have been carried out to detect them in the sub-TeV energy band. Up to now, only upper 
limits are provided by H.E.S.S., VERITAS and MAGIC. Expectation for MAGIC to detect GRB is due to 
its low energy threshold which increases the acceptance circle at $z=0.2$--0.5, and for its technical 
capability of fast  re-pointing  the telescope following a GRB alert.
%
%
\subsection{Galactic sources} 
\begin{figure}
\resizebox{\hsize}{!}{\includegraphics[clip=true]{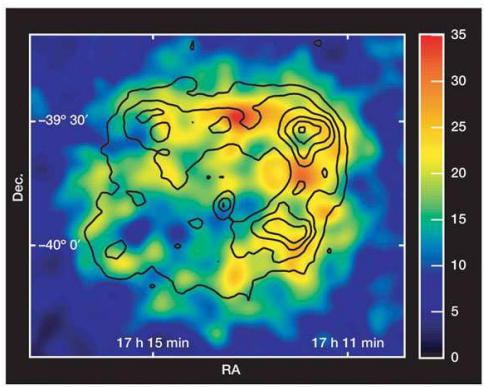}}
\caption{\footnotesize
$\gamma$-ray image of the SNR RX~J1713.7$-$3946 obtained with the H.E.S.S. telescopes. ASCA
1--3 keV black contours are superimposed to the H.E.S.S. $\gamma$-ray counts map.
\cite{Aharonian2004}.
}
\label{fig:rxj1713}
\end{figure}

%
%
The majority of TeV sources are Galactic objects confined at low latitude, at a few kiloparsec  distance,  
and in some case extended. Galactic source classes include: shell type supernova remnants
(see Figure~\ref{fig:rxj1713}),   
pulsar wind nebulae, binary systems, the Galactic Center, and unidentified sources 
\citep{Hinton2009}.
The TeV SNR emission is linked to the cosmic ray paradigm and allows the 
unambiguous identification of the sites where cosmic particles are accelerated and diffused. 
The first-order acceleration process is a very efficient mechanism to accelerate particles up to 
$10^{15}$ eV to produce $\gamma$-rays by inverse Compton (electrons) or by interaction 
with nuclei (hadrons).

PWNs constitute a relatively large (about 20 objects) important class of Galactic sources 
in which the TeV emission seems to be produced by inverse Compton of particles accelerated 
in the pulsar magnetosphere. 
One of these sources is the Crab Nebula. An important recent result is the discovery by MAGIC of a 
pulsed component of the high $\gamma$-ray photons above 25 GeV \citep{Aliu2008:Crab}. 
Currently, the Crab is used as standard candle for the Cherenkov telescopes calibrations. 
The recent discovery by AGILE \citep{Tavani2010:Crab} and its confirmation by 
Fermi/LAT \citep{Buehler2010:Crab} of an unexpected  flare at energy above 100 MeV from a region
consistent with the Crab Nebula itself challenges the current theoretical models, opening new scenarios, and
arises severe warnings on the past and future calibration procedures.

H.E.S.S. discovered TeV emission from three binary systems, i.e.:  PSR B~1259$-$63
\citep{Aharonian2005:PSRB1259}, LS~5039 \citep{Aharonian2005:LS5039}, and 
LS~I~$+$61~303 \citep{Albert2006:LSI61303}. 

The stellar-mass black-hole binary Cygnus X--1 has been detected by MAGIC only in one episodic flare and it is
unclear whether this emission shows a periodicity or not \citep{Albert2007:cygx1}. 
The flux variability exhibited by this class of objects
could be the reason why the TeV emission by other binaries detected in the past by other experiments has not been
confirmed up to now.  
In the Galactic Center region the position of the source H.E.S.S. unidentified source HESS~J1745$-$290
is, within the error circle, coincident with the position of the supermassive black hole Sgr A$^{\star}$, even if
other alternative identifications cannot be excluded, because of the high number of sources in the Galactic Center  
\citep{Tsuchiya2004:sgra, Acero2010MNRAS:sgra}.
Moreover, H.E.S.S. performed a survey of the Milky Way. As result of this survey, together with serendipitous discoveries 
in several fields of known sources, a high number of new sources have been discovered. The majority of them do 
not have any counterpart at other wavelengths. More than 20 sources remain unidentified, many of them are point-like
Galactic sources, with the exception of HESS~J1303-6 which appears to be extended \citep{Aharonian2005:unid}.
%
%
\subsection{Fundamental Physics: Search for dark matter annihilation and for Lorentz invariance violation.} 
%
%
Very high-energy $\gamma$-ray photons could be tracers of dark matter because they are the result
of the elastic scattering with the matter or of the self-annihilation of the weakly-interacting massive particles (WIMPs),
the most popular component of the dark matter. 
The WIMPs mass range lies in the $\gamma$-ray energy band, up to a few hundred of GeV, making Cherenkov 
Telescopes potential detectors of this phenomenon. 

The extreme energy band of the TeV Astronomy, several orders of magnitude from the optical or infrared astronomy, 
can be used to verify the  Lorentz-invariance violation of the speed of light induced by the Quantum Gravity effect. 
The experimental way is to measure a time lag between low (optical/infrared) and high energy (TeV) photons emitted 
in a short pulse by  a distant  cosmological source. Up to now only preliminary results were obtained
\citep{Aharonian2008:2155,Albert2008:mkn501,Martinez2009:Lorentz}.
Future Cherenkov Compton Arrays, with their improved sensitivity,  will be more suitable for these kinds of 
investigations in the fields of fundamental physics and cosmology. 
%
%
\section{The CTA, as future, open, big infrastructure for the TeV astronomy}
%
%
\subsection{The CTA array} 
%
%
Thanks to several X-ray and $\gamma$-ray observatories already in orbit (e.g., Chandra, XMM-Newton, 
Swift, Fermi, AGILE) and to many ground- and space-based optical and infrared telescopes (e.g., REM, HST, 
SPITZER), it will be possible  to study the Universe, for the first time, all over the electromagnetic spectrum almost simultaneously. 
\begin{figure}
\resizebox{\hsize}{!}{\includegraphics[clip=true]{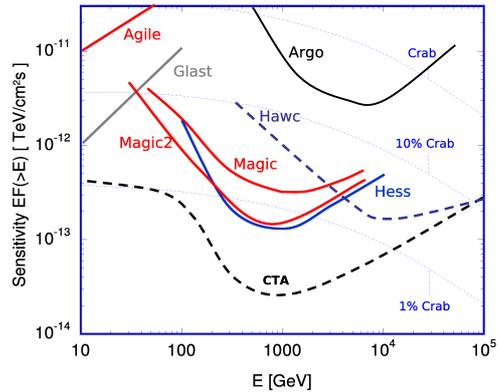}}
\caption{\footnotesize
Foreseen CTA sensitivity in the energy interval 0.01--100 TeV compared with that of the  
present Cherenkov Arrays. \cite{DeAngelis2008}.
}
\label{fig:cta_sens}
\end{figure}
In such a scenario, a new generation of ground-based VHE $\gamma$-ray instruments is needed in 
order to significantly improve the sensitivity, the observed energy band, the field of view, and to reduce 
the integration time. 
The next generation Cherenkov Telescope Array will significantly improve the performance of the 
current VHE telescopes in the following lines: 
\begin{itemize}
	\item Gain one order of magnitude in sensitivity at 1 TeV (1 mCrab), increasing the number of the detected TeV sources of a factor of 10--100 and allowing accurate study of fundamental physics phenomena.
	\item Reach 10--20 GeV as the lower energy threshold to link the TeV with the GeV spectra as measured from space missions. This is particularly important for AGN, EBL studies and GRB detection.
	\item Reach 100--200 TeV as the upper  energy threshold, in order to open the possibility to investigate the sites where the cosmic rays are accelerated and diffused, at about the ``knee'' energy of CR spectrum.
	\item Improve  the angular resolution, to pursue morphological studies of extended sources, also in correlation with other wavelengths, and to restrict the error boxes of the unidentified sources, allowing an easier identification.
	\item Enlarge the field of view, to carry-out faster surveys, increasing the number of serendipitous discoveries,  and to perform simultaneous observations of extended sources and wide sky zones for dark matter searches.   
\end{itemize}
\begin{figure*}[ht!]
\resizebox{\hsize}{!}{\includegraphics[clip=true]{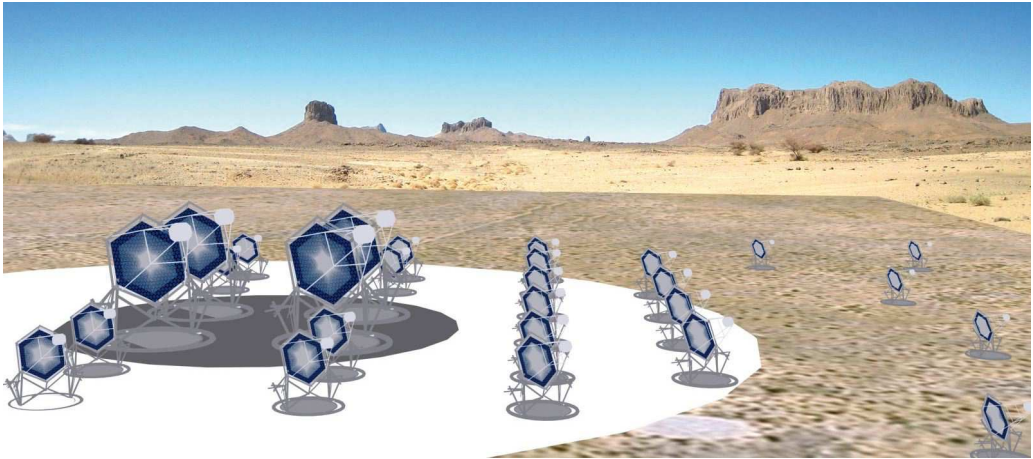}}
\caption{\footnotesize
Artist's view of the CTA, with three different telescopes types covering the overlapping energy ranges, and area 
coverage which increases with increasing $\gamma$-ray energy. \cite{CTA:design:concept:2010}
}
\label{fig:cta_array}
\end{figure*}

The international scientific community operating in the TeV astronomy  by now considers the idea mature to go
beyond experiments such as  H.E.S.S., VERITAS, and MAGIC towards a new, much larger, Cherenkov Telescope 
Array characterized by a very large number of telescopes, with new technological improvements both for mirrors and for
sensors, and to open this Facility to all the scientists as an Observatory on the basis of the scientific merit.  
Figure \ref{fig:cta_sens} shows the sensitivity foreseen for CTA, spanning five decades in energy, compared with 
the current experiments.

The CTA better sensitivity level and the larger energy interval can be reached by an array composed of 
many telescopes of three different types, each type tuned to a particular energy band:
\begin{itemize}
	\item the weak light produced by $\gamma$-rays between 10 GeV and 100 GeV energy will be detected by 4 telescopes of 24 m dishes, evolution of the MAGIC telescopes (the large-size telescopes, LSTs);
	\item following the HESS approach, the central energies (100 GeV - 10 TeV) will be covered by several tens of 12 m class telescopes (the medium-size telescopes, MSTs);
	\item for the extreme energy range above 10 TeV, a large number of smaller light collector, 3-8 m, spread over a fiducial area of 4--10 km$^2$ is suitable for these strong but rare events (the small-size telescopes, SSTs).
\end{itemize}
Figure \ref{fig:cta_array} show an artist's view of the CTA, with three different telescopes types.

%
%
\subsection{The Cherenkov Telescope optical design} 
\begin{figure}[ht!]
\resizebox{\hsize}{!}{
\includegraphics[clip=true]{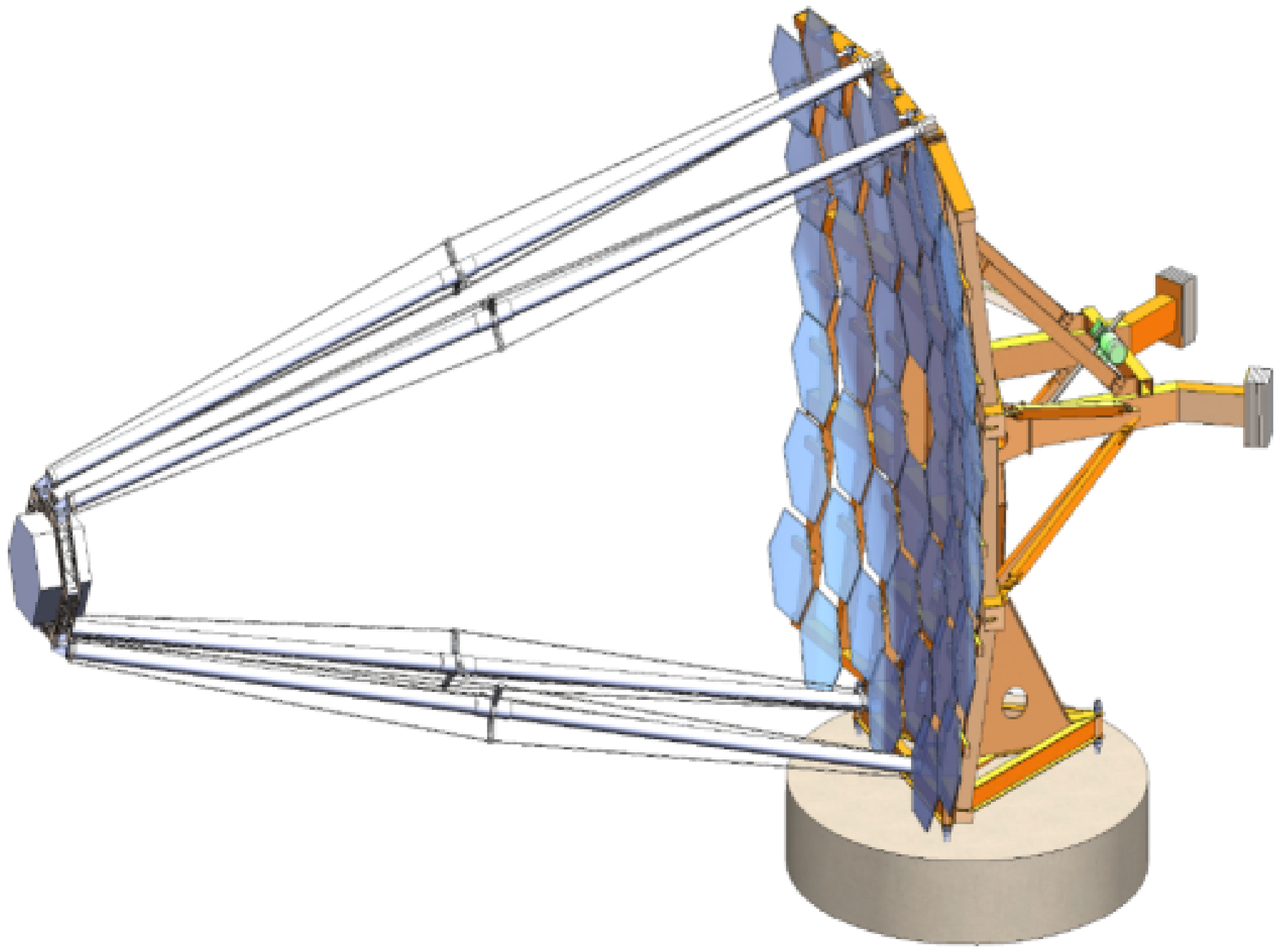}
}
\caption{\footnotesize
The Davies-Cotton design. \cite{CTA:design:concept:2010}
}
\label{fig:dc}
\end{figure}

%
%
Traditional Cherenkov telescopes follow the single-reflection Davies-Cotton like (DC) design. 
The DC  design consists of a large mirror, which concentrates the Cherenkov light flash on a segmented 
camera. In the current Cherenkov telescopes, each camera pixel is a single photo-multiplier (PMT) capable to 
detect very fast signals at the nanosecond level. Figure \ref{fig:dc} shows a Cherenkov telescope model proposed for the 
CTA SST sub-array \citep{CTA:design:concept:2010}. 

\begin{figure}[ht!]
\resizebox{\hsize}{!}{
\includegraphics[clip=true]{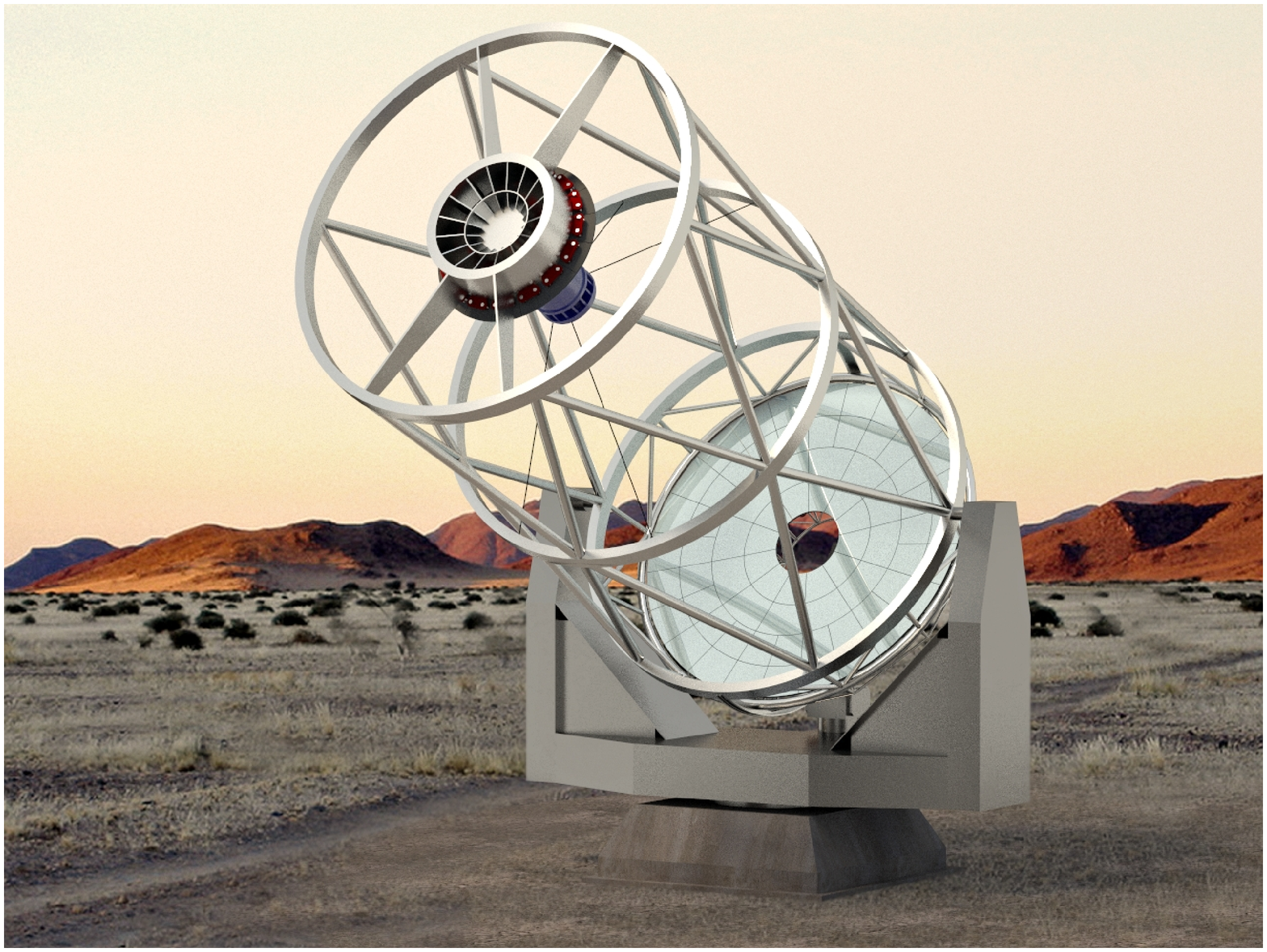}
}
\caption{\footnotesize
The Schwarzschild-Couder design. \cite{CTA:design:concept:2010}.
}
\label{fig:sc}
\end{figure}

An alternative design to the single-reflection telescope has been studied for the Advanced Gamma-ray 
Imaging System (AGIS) array \citep{Vassiliev2007:AGIS}, consisting of 36 dual-mirror Schwarzschild-Couder 
(SC) telescopes. One of the advantages of using the SC design with respect to the DC is the higher angular 
resolution reachable by each single telescope. From the point of view of feasibility and costs of the two designs,  
single-mirror telescopes are easier to manufacture, because of their larger curvature radius, but the large focal 
length introduces a complex and heavy camera placed at large distance from the dish structure.
For the same collecting area, dual-mirror telescopes have a more compact structure and a smaller camera,
with the possibility of using multi-pixel sensors. Unfortunately, because of the shorter focal length, mirrors have stronger
curvatures and, therefore, they can be manufactured only by means of more complex technologies. All in all, 
single-mirror telescopes are more affordable for the mirror costs, while dual-mirror telescopes are more affordable  
for the camera and structure costs. 
Figure \ref{fig:sc} shows a dual-mirror telescope proposed by INAF scientists (Canestrari et al, 2010, 
private communication) for the CTA SST sub-array, as an alternative design to the one shown in Figure \ref{fig:dc}. 
Within the first year of the Preparatory Phase, the CTA Consortium will do the final choice between the two 
alternative models.
%
%
%
%
\section{The CTA consortium and the Italian participation}
%
%
The goal of CTA Consortium is to build a ground-based facility capable to measure electromagnetic radiation from sky 
at energies between 10 GeV and hundreds of TeV with unprecedented sensitivity, energy interval, and angular resolution.
Unlike the present arrays (H.E.S.S., MAGIC, VERITAS) conceived as experiments, CTA will be operated as an Observatory 
open to scientists of the world scientific communities including astronomy and astrophysics, cosmology, astroparticle 
physics, and particle physics. 
CTA will explore the sky participating to multi-wavelength studies with radio, infrared, optical, X-rays and 
$\gamma$-rays facilities extending to very high energies the electromagnetic spectrum to probe the extreme processes 
in the Universe.

The CTA Consortium is composed of more than 100 scientific Institutions belonging to 22 countries :
14  from the European Union  (Bulgaria, Czech Republic, Denmark, Finland, France, Germany, Greece, Italy, Ireland, 
Holland, Poland, Spain, Sweden  and United Kingdom), 3 European (Armenia, Croatia and  Switzerland) and 
5 extra-european (Argentine, Japan, Namibia, South Africa e United States of America). 
{\it In fieri} is the participation from Brasil, India, and Slovenia.
Since 2008, CTA is in the road-map of the European Strategy Forum on Research and Infrastructure (ESFRI).
In 2009 CTA has been evaluated by ASPERA as one of the seven most important European projects for the 
Astro-particle physics. In 2010, the European Commission approved the Preparatory Phase CTA Programme in the 
framework of the FP7-INFRACTURES-2010-1 and funded it for 5.2 MEuro. The Preparatory Phase starts on 
October 1th, 2010 and will end on September 30th, 2013. 
In 2013 it is foreseen the start of the construction phase, while in 2018 all the infrastructure will be completed. 
The scientific activity of CTA can be carried out before the completion of the total array, starting from when it is only 
10\% complete.

The Italian participation to the CTA Consortium includes INAF (scientific Institutes of Bologna, Catania, Milano, 
Padova, Palermo, Roma, Torino, and TNG) and the Universities of Padova, Siena, and Udine.
The Italian contribution is in the mirror technology and production, sensors (Multi Anode PMTs and SiPM) 
electronics, design of the SST, data handling, and governance. INAF is also responsible 
of the industrial procurement activity.


\begin{acknowledgements}
Authors thank INAF Dept. of Projects for the financial and operative support to the participation to the CTA Preparatory Phase. B.S. thanks the SOC of the 54th SAIT Meeting for the invitation.
\end{acknowledgements}

\bibliographystyle{aa}

\end{document}